\documentclass[12pt,a4paper]{article}
\usepackage{booktabs}
\usepackage{subfig}
\usepackage{placeins}
\usepackage{mathtools}
\usepackage{ifthen} 
\newboolean{pdflatex}
\setboolean{pdflatex}{true} 

\newboolean{articletitles}
\setboolean{articletitles}{true} 

\newboolean{uprightparticles}
\setboolean{uprightparticles}{false} 

\newboolean{inbibliography}
\setboolean{inbibliography}{true} 

\usepackage{graphicx}
\usepackage{nicefrac} 

\usepackage{microtype}
\usepackage{lineno}  
\usepackage{xspace} 
\usepackage{caption} 

\usepackage{graphicx}  
\usepackage{color}
\usepackage{colortbl}
\graphicspath{{./figs/}} 

\usepackage{amsmath} 
\usepackage{amssymb}
\usepackage{amsfonts}
\usepackage{upgreek} 

\newcommand*\patchAmsMathEnvironmentForLineno[1]{%
\expandafter\let\csname old#1\expandafter\endcsname\csname #1\endcsname
\expandafter\let\csname oldend#1\expandafter\endcsname\csname
end#1\endcsname
 \renewenvironment{#1}%
   {\linenomath\csname old#1\endcsname}%
   {\csname oldend#1\endcsname\endlinenomath}%
}
\newcommand*\patchBothAmsMathEnvironmentsForLineno[1]{%
  \patchAmsMathEnvironmentForLineno{#1}%
  \patchAmsMathEnvironmentForLineno{#1*}%
}
\AtBeginDocument{%
\patchBothAmsMathEnvironmentsForLineno{equation}%
\patchBothAmsMathEnvironmentsForLineno{align}%
\patchBothAmsMathEnvironmentsForLineno{flalign}%
\patchBothAmsMathEnvironmentsForLineno{alignat}%
\patchBothAmsMathEnvironmentsForLineno{gather}%
\patchBothAmsMathEnvironmentsForLineno{multline}%
}

\usepackage{hyperref}    
\usepackage[all]{hypcap} 

\def\lhcb {\mbox{LHCb}\xspace}

\ifthenelse{\boolean{uprightparticles}}%
{
 
 \def\Pgamma      {\ensuremath{\upgamma}\xspace}

 \def\Ppi         {\ensuremath{\uppi}\xspace}

 \def\PDelta      {\ensuremath{\Delta}\xspace}                 
 \def\PXi      {\ensuremath{\Xi}\xspace}                 
 \def\PLambda      {\ensuremath{\Lambda}\xspace}                 
 \def\PSigma      {\ensuremath{\Sigma}\xspace}                 
 \def\POmega      {\ensuremath{\Omega}\xspace}                 
 \def\PUpsilon      {\ensuremath{\Upsilon}\xspace}                 
 
 \def\PB      {\ensuremath{\mathrm{B}}\xspace}                 
                  
 \def\PD      {\ensuremath{\mathrm{D}}\xspace}

 \def\PK      {\ensuremath{\mathrm{K}}\xspace}

 \def\PW      {\ensuremath{\mathrm{W}}\xspace}

 \def\Pb      {\ensuremath{\mathrm{b}}\xspace}                 
 \def\Pc      {\ensuremath{\mathrm{c}}\xspace}

 \def\Pi      {\ensuremath{\mathrm{i}}\xspace}

 \def\Ps      {\ensuremath{\mathrm{s}}\xspace}

}
{
 
 \def\Pgamma      {\ensuremath{\gamma}\xspace}

 \def\Ppi         {\ensuremath{\pi}\xspace}

 \mathchardef\PDelta="7101
 \mathchardef\PXi="7104
 \mathchardef\PLambda="7103
 \mathchardef\PSigma="7106
 \mathchardef\POmega="710A
 \mathchardef\PUpsilon="7107
                  
 \def\PB      {\ensuremath{B}\xspace}                 
                  
 \def\PD      {\ensuremath{D}\xspace}

 \def\PK      {\ensuremath{K}\xspace}

 \def\PW      {\ensuremath{W}\xspace}

 \def\Pb      {\ensuremath{b}\xspace}                 
 \def\Pc      {\ensuremath{c}\xspace}

 \def\Pi      {\ensuremath{i}\xspace}

 \def\Ps      {\ensuremath{s}\xspace}

}

\def\g      {{\ensuremath{\Pgamma}}\xspace}

\def\W      {{\ensuremath{\PW}}\xspace}

\def\squark    {{\ensuremath{\Ps}}\xspace}

\def\cquark    {{\ensuremath{\Pc}}\xspace}

\def\bquark    {{\ensuremath{\Pb}}\xspace}

\def\pion   {{\ensuremath{\Ppi}}\xspace}
\def\piz    {{\ensuremath{\pion^0}}\xspace}

\def\pip    {{\ensuremath{\pion^+}}\xspace}
\def\pim    {{\ensuremath{\pion^-}}\xspace}
\def\pipm   {{\ensuremath{\pion^\pm}}\xspace}
\def\pimp   {{\ensuremath{\pion^\mp}}\xspace}

\def\kaon    {{\ensuremath{\PK}}\xspace}
  \def\Kbar    {{\kern 0.2em\overline{\kern -0.2em \PK}{}}\xspace}

\def\Kp      {{\ensuremath{\kaon^+}}\xspace}

\def\Kpm     {{\ensuremath{\kaon^\pm}}\xspace}

\def\Kstar   {{\ensuremath{\kaon^*}}\xspace}

  \def\Dbar    {{\kern 0.2em\overline{\kern -0.2em \PD}{}}\xspace}
\def\D       {{\ensuremath{\PD}}\xspace}

\def\Dz      {{\ensuremath{\D^0}}\xspace}

\def\Dpm     {{\ensuremath{\D^\pm}}\xspace}

\def\B       {{\ensuremath{\PB}}\xspace}
\def\Bbar    {{\ensuremath{\kern 0.18em\overline{\kern -0.18em \PB}{}}}\xspace}

\def\Bz      {{\ensuremath{\B^0}}\xspace}

\def\Bu      {{\ensuremath{\B^+}}\xspace}
\def\Bub     {{\ensuremath{\B^-}}\xspace}
\def\Bp      {{\ensuremath{\Bu}}\xspace}
\def\Bm      {{\ensuremath{\Bub}}\xspace}
\def\Bpm     {{\ensuremath{\B^\pm}}\xspace}

  \def\Y#1S{\ensuremath{\PUpsilon{(#1S)}}\xspace}

\def\Lbar        {{\ensuremath{\kern 0.1em\overline{\kern -0.1em\PLambda}}}\xspace}

\newcommand{\decay}[2]{\ensuremath{#1\!\to #2}\xspace}         

\def\to                 {\ensuremath{\rightarrow}\xspace}

\def\CP                {{\ensuremath{C\!P}}\xspace}

\def\AT#1     {\ensuremath{A_{\mathrm{T}}^{#1}}\xspace}           
\def\btosgam  {\decay{\bquark}{\squark \g}}

\def\C#1      {\ensuremath{\mathcal{C}_{#1}}\xspace}                       
\def\Cp#1     {\ensuremath{\mathcal{C}_{#1}^{'}}\xspace}                    
\def\Ceff#1   {\ensuremath{\mathcal{C}_{#1}^{\mathrm{(eff)}}}\xspace}        
\def\Cpeff#1  {\ensuremath{\mathcal{C}_{#1}^{'\mathrm{(eff)}}}\xspace}       
\def\Ope#1    {\ensuremath{\mathcal{O}_{#1}}\xspace}                       
\def\Opep#1   {\ensuremath{\mathcal{O}_{#1}^{'}}\xspace}                    

\newcommand{\tev}{\ifthenelse{\boolean{inbibliography}}{\ensuremath{~T\kern -0.05em eV}\xspace}{\ensuremath{\mathrm{\,Te\kern -0.1em V}}}\xspace}
\newcommand{\gev}{\ensuremath{\mathrm{\,Ge\kern -0.1em V}}\xspace}
\newcommand{\mev}{\ensuremath{\mathrm{\,Me\kern -0.1em V}}\xspace}
\newcommand{\kev}{\ensuremath{\mathrm{\,ke\kern -0.1em V}}\xspace}
\newcommand{\ev}{\ensuremath{\mathrm{\,e\kern -0.1em V}}\xspace}
\newcommand{\gevc}{\ensuremath{{\mathrm{\,Ge\kern -0.1em V\!/}c}}\xspace}
\newcommand{\mevc}{\ensuremath{{\mathrm{\,Me\kern -0.1em V\!/}c}}\xspace}
\newcommand{\gevcc}{\ensuremath{{\mathrm{\,Ge\kern -0.1em V\!/}c^2}}\xspace}
\newcommand{\gevgevcccc}{\ensuremath{{\mathrm{\,Ge\kern -0.1em V^2\!/}c^4}}\xspace}
\newcommand{\mevcc}{\ensuremath{{\mathrm{\,Me\kern -0.1em V\!/}c^2}}\xspace}

\def\mum  {\ensuremath{{\,\upmu\rm m}}\xspace}

\def\invfb   {\ensuremath{\mbox{\,fb}^{-1}}\xspace}

\newcommand{\chisq}{\ensuremath{\chi^2}\xspace}

\def\gsim{{~\raise.15em\hbox{$>$}\kern-.85em
          \lower.35em\hbox{$\sim$}~}\xspace}
\def\lsim{{~\raise.15em\hbox{$<$}\kern-.85em
          \lower.35em\hbox{$\sim$}~}\xspace}

\def\PDF {PDF\xspace}

\def\pt         {\mbox{$p_{\rm T}$}\xspace}

\def\evtgen     {\mbox{\textsc{EvtGen}}\xspace}

\def\geant      {\mbox{\textsc{Geant4}}\xspace}

\def\photos     {\mbox{\textsc{Photos}}\xspace}

\def\pythia     {\mbox{\textsc{Pythia}}\xspace}

\def\tell1  {TELL1\xspace}
\def\ukl1   {UKL1\xspace}

\usepackage{cite} 
\usepackage{mciteplus}

\begin{document}

\renewcommand{\thefootnote}{\fnsymbol{footnote}}
\setcounter{footnote}{1}


\begin{titlepage}
\pagenumbering{roman}

\vspace*{-1.5cm}
\centerline{\large EUROPEAN ORGANIZATION FOR NUCLEAR RESEARCH (CERN)}
\vspace*{1.5cm}
\hspace*{-0.5cm}
\begin{tabular*}{\linewidth}{lc@{\extracolsep{\fill}}r}
\ifthenelse{\boolean{pdflatex}}
{\vspace*{-2.7cm}\mbox{\!\!\!\includegraphics[width=.14\textwidth]{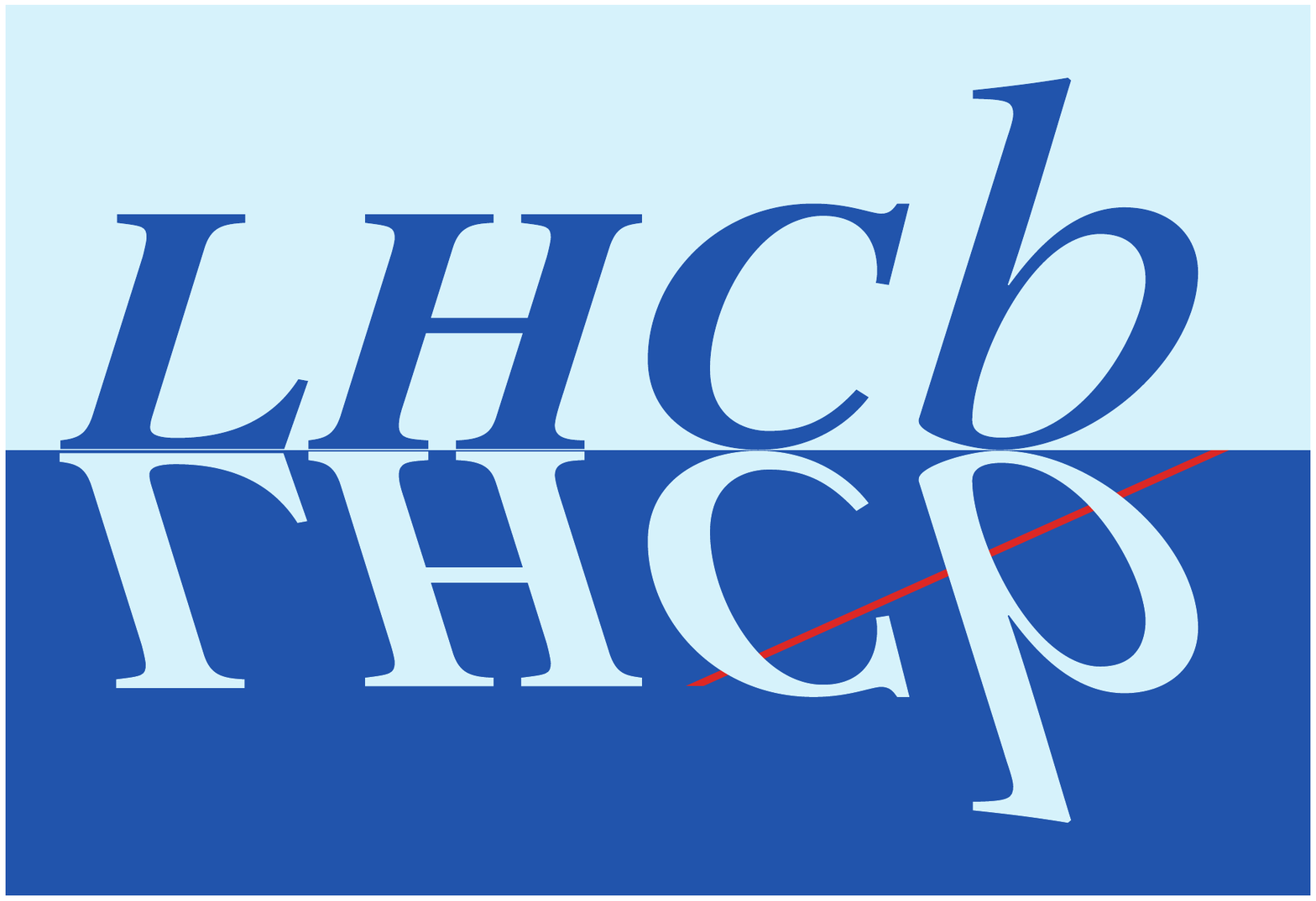}} & &}%
{\vspace*{-1.2cm}\mbox{\!\!\!\includegraphics[width=.12\textwidth]{lhcb-logo.eps}} & &}%
\\
 & & CERN-PH-EP-2014-026 \\  
 & & LHCb-PAPER-2014-001 \\  
 & & 27 February 2014\\
 & & \\
\end{tabular*}

\vspace*{3.0cm}

{\bf\boldmath\huge
\begin{center}
    Observation of photon polarization in the \btosgam transition
\end{center}
}

\vspace*{1.0cm}

\begin{center}
The LHCb collaboration\footnote{Authors are listed on the following pages.}
\end{center}

\vspace{\fill}

\begin{abstract}
  \noindent This Letter presents a study of the flavor-changing neutral current radiative \decay{\Bpm}{\Kpm\pimp\pipm\g} decays performed using data collected in proton-proton collisions with the \lhcb detector at $7$ and $8\,$TeV center-of-mass energies. 
In this sample, corresponding to an integrated luminosity of $3\,\invfb$, nearly $14\,000$ signal events are reconstructed and selected, containing all possible intermediate resonances with a $\Kpm\pimp\pipm$ final state in the $[1.1, 1.9]$\,\gevcc mass range. 
The distribution of the angle of the photon direction with respect to the plane defined by the final-state hadrons in their rest frame is studied in intervals of $\Kpm\pimp\pipm$ mass and the asymmetry between the number of signal events found on each side of the plane is obtained.
The first direct observation of the photon polarization in the \btosgam transition is reported with a significance of $5.2\,\sigma$.

\end{abstract}

\vspace*{1.0cm}

\begin{center}
  Published in Phys.~Rev.~Lett. 112, 161801 (2014)
\end{center}

\vspace{\fill}

{\footnotesize 
\centerline{\copyright~CERN on behalf of the \lhcb collaboration, license \href{http://creativecommons.org/licenses/by/3.0/}{CC-BY-3.0}.}}
\vspace*{2mm}

\end{titlepage}


\newpage
\setcounter{page}{2}
\mbox{~}
\newpage

\centerline{\large\bf LHCb collaboration}
\begin{flushleft}
\small
R.~Aaij$^{41}$, 
B.~Adeva$^{37}$, 
M.~Adinolfi$^{46}$, 
A.~Affolder$^{52}$, 
Z.~Ajaltouni$^{5}$, 
J.~Albrecht$^{9}$, 
F.~Alessio$^{38}$, 
M.~Alexander$^{51}$, 
S.~Ali$^{41}$, 
G.~Alkhazov$^{30}$
% ...

{\footnotesize \it
% ...
$ ^{5}$Clermont Universit\'{e}, Universit\'{e} Blaise Pascal, CNRS/IN2P3, LPC, Clermont-Ferrand, France\\
% ...
$ ^{9}$Fakult\"{a}t Physik, Technische Universit\"{a}t Dortmund, Dortmund, Germany\\
% ...
$ ^{30}$Petersburg Nuclear Physics Institute (PNPI), Gatchina, Russia\\
% ...
$ ^{37}$Universidad de Santiago de Compostela, Santiago de Compostela, Spain\\
$ ^{38}$European Organization for Nuclear Research (CERN), Geneva, Switzerland\\
% ...
$ ^{41}$Nikhef National Institute for Subatomic Physics, Amsterdam, The Netherlands\\
% ...
$ ^{46}$H.H. Wills Physics Laboratory, University of Bristol, Bristol, United Kingdom\\
% ...
$ ^{51}$School of Physics and Astronomy, University of Glasgow, Glasgow, United Kingdom\\
$ ^{52}$Oliver Lodge Laboratory, University of Liverpool, Liverpool, United Kingdom\\
% ...
}
\end{flushleft}

\cleardoublepage


\renewcommand{\thefootnote}{\arabic{footnote}}
\setcounter{footnote}{0}
\cleardoublepage

\setcounter{tocdepth}{2}


\pagestyle{plain} 
\setcounter{page}{1}
\pagenumbering{arabic}


\newcommand{\mysection}[1]{\section[#1]{\boldmath #1}}
\newcommand{\mysubsection}[1]{\subsection[#1]{\boldmath #1}}
\newcommand{\mysubsubsection}[1]{\subsubsection[#1]{\boldmath #1}}

\newcommand{\Kone}[1]{\ensuremath{\kaon_1(#1)}}
\newcommand{\Konep}[1]{\ensuremath{\kaon_1(#1)^+}}
\newcommand{\Konem}[1]{\ensuremath{\kaon_1(#1)^-}}
\newcommand{\Konez}[1]{\ensuremath{\kaon_1(#1)^0}}
\newcommand{\Kst}[2]{\ensuremath{\Kstar_{#1}(#2)}}
\newcommand{\Kstp}[2]{\ensuremath{\kaon^{*}_{#1}(#2)^+}}
\newcommand{\Kstm}[2]{\ensuremath{\Kstar_{#1}(#2)^-}}
\newcommand{\Kstz}[2]{\ensuremath{\Kstar_{#1}(#2)^0}}
\newcommand{\Ktwo}[1]{\ensuremath{\kaon_1(#1)}}
\newcommand{\diff}[1]{\operatorname{d}\!#1}

\def \Kres{\ensuremath{\kaon_{\text{res}}}\xspace}
\def \kpp{\ensuremath{\kaon\pion\pion}\xspace}
\def \Kpipi{\ensuremath{\kaon^+\pion^-\pion^+}\xspace}
\def \pipipi{\ensuremath{\pion\pion\pion}\xspace}
\def \Kpipig{\ensuremath{\Kp\pim\pip\g}\xspace}
\def \BtoKpipiEta{\decay{\Bp}{\Kp\pim\pip\eta}}
\def \BtoKpipig{\decay{\Bp}{\Kp\pim\pip\g}}
\def \BtoKrestoKpipig{\decay{\B^+}{\decay{K^+_{\text{res}}{\g}}{\Kpipi\g}}}
\def \BtoKonetoKpipig{\decay{\Bp}{\decay{K_1^+{\g}}{\Kp\pim\pip\g}}}
\def \Btopipipig{\decay{\Bp}{\pip\pim\pip\g}} 
\def \BztoKresg{\decay{\Bz}{\decay{K_1^0 \g}{\Kp\pim\piz} \g}}
\newcommand{\BtoKoneRes}[1]{\ensuremath{\decay{\Bp}{\kaon_1(#1)^{+}{\g}}}\xspace}
\newcommand{\BtoKstRes}[2]{\ensuremath{\decay{\Bp}{\kaon_{#1}(#2)^{+}{\g}}}\xspace}
\newcommand{\BtoKststRes}[2]{\ensuremath{\decay{\Bp}{\kaon^*_{#1}(#2)^{+}{\g}}}\xspace}

\def \Uplus{\ensuremath{U^+}}
\def \Uminus{\ensuremath{U^-}}
\def \Upm{\ensuremath{U^{\pm}}\xspace}
\def \Dplus{\ensuremath{D^+}}
\def \Dminus{\ensuremath{D^-}}
\def \Dpm{\ensuremath{D^{\pm}}\xspace}
\def \CB{\ensuremath{C\!B}}
\def \costheta{\ensuremath{\cos\theta}\xspace}
\newcommand{\asym}[1]{\ensuremath{\mathcal{A}^{#1}_{\text{ud}}}\xspace}
\newcommand{\aup}{\ensuremath{\mathcal{A}_{\text{ud}}}\xspace}
\newcommand{\acp}{\ensuremath{\mathcal{A}_{\CP}}\xspace}
\newcommand{\acpRaw}{\ensuremath{\mathcal{A}_{\CP}^{\text{raw}}}\xspace}
\newcommand{\Dasym}{\ensuremath{\mathcal{D}}\xspace}
\def \ap {\ensuremath{\mathcal{A}_{\text{P}}}\xspace}
\def \ad {\ensuremath{\mathcal{A}_{\text{D}}}\xspace}

\newcommand{\Nsig}[1]{\ensuremath{N_{\text{signal}}^{#1}}\xspace}
\newcommand{\Nbkg}[1]{\ensuremath{N_{\text{bkg}}^{#1}}\xspace}

\def \fMissingPi{\ensuremath{f_{{\text{miss-}\pion}}}\xspace}
\def \fPartial{\ensuremath{f_{\text{partial}}}\xspace}
\def \cPartial{\ensuremath{c_{\text{partial}}}\xspace}
\def \pPartial{\ensuremath{p_{\text{partial}}}\xspace}
\def \NBkgPlus{\ensuremath{N^{+}_{\text{bkg}}}\xspace}
\def \NBkgMinus{\ensuremath{N^{-}_{\text{bkg}}}\xspace}
\def \fMissingPiPlus{\ensuremath{f^{+}_{{\text{miss-}\pion}}}\xspace}
\def \fMissingPiMinus{\ensuremath{f^{-}_{{\text{miss-}\pion}}}\xspace}
\def \fPartialMinus{\ensuremath{f^{-}_{\text{partial}}}\xspace}
\def \fPartialPlus{\ensuremath{f^{+}_{\text{partial}}}\xspace}
\def \NBkgPlusUp{\ensuremath{N^{+ \text{,up}}_{\text{bkg}}}\xspace}
\def \NBkgPlusDown{\ensuremath{N^{+ \text{,down}}_{\text{bkg}}}\xspace}
\def \NBkgMinusUp{\ensuremath{N^{- \text{,up}}_{\text{bkg}}}\xspace}
\def \NBkgMinusDown{\ensuremath{N^{- \text{,down}}_{\text{bkg}}}\xspace}
\def \fMissingPiPlusUp{\ensuremath{f^{+ \text{,up}}_{{\text{miss-}\pion}}}\xspace}
\def \fMissingPiPlusDown{\ensuremath{f^{+ \text{,down}}_{{\text{miss-}\pion}}}\xspace}
\def \fMissingPiMinusUp{\ensuremath{f^{- \text{,up}}_{{\text{miss-}\pion}}}\xspace}
\def \fMissingPiMinusDown{\ensuremath{f^{- \text{,down}}_{{\text{miss-}\pion}}}\xspace}
\def \fPartialMinusDown{\ensuremath{f^{- \text{,down}}_{\text{partial}}}\xspace}
\def \fPartialMinusUp{\ensuremath{f^{- \text{,up}}_{\text{partial}}}\xspace}
\def \fPartialPlusDown{\ensuremath{f^{+ \text{,down}}_{\text{partial}}}\xspace}
\def \fPartialPlusUp{\ensuremath{f^{+ \text{,up}}_{\text{partial}}}\xspace}
\def \tauPlus{\ensuremath{\tau^{+}}\xspace}
\def \tauMinus{\ensuremath{\tau^{-}}\xspace}
\def\Kresbar{\ensuremath{\Kbar_{\text{res}}}\xspace}
\def\Kresbarpol{\ensuremath{\Kbar_{\text{res,pol}}}\xspace}
\def\BKresgammaPPPgamma{\ensuremath{\decay{\Bp}{\decay{\Kres^{+}\gamma}{P_1P_2P_3\g}}}\xspace}
\def\BPPPgamma{\ensuremath{\decay{\Bbar}{P_1P_2P_3\g}}\xspace}
\def\BKresgamma{\ensuremath{\decay{\Bbar}{\Kresbar\gamma}}\xspace}
\def\BKresgammapol{\ensuremath{\decay{\Bbar}{\Kresbarpol\gamma_{\text{pol}}}}\xspace}
\def\KresPPP{\ensuremath{\decay{\Kresbar}{P_1P_2_P3}}\xspace}
\def\KrespolPPP{\ensuremath{\decay{\Kresbarpol}{P_1P_2P_3}}\xspace}

\def\costtheta{\ensuremath{\cos\tilde{\theta}}\xspace}
\def\coshtheta{\ensuremath{\cos\hat{{\theta}}}\xspace}

\newpage

The Standard Model (SM) predicts that the photon emitted from the electroweak penguin loop in \btosgam transitions is predominantly left-handed, since the recoiling \squark quark that couples to a \W boson is left-handed.
This implies maximal parity violation up to small corrections of the order $m_{\squark}/m_{\bquark}$.
While the measured inclusive \btosgam rate~\cite{PDG2012} agrees with the SM calculations, no direct evidence exists for a nonzero photon polarization in this type of decays.
Several extensions of the SM~\cite{Pati:1974yy,*Mohapatra:1974gc,*Mohapatra:1974hk,*Senjanovic:1975rk,*Senjanovic:1978ev,*Mohapatra:1980yp,*Lim:1981kv,*Everett:2001yy}, compatible with all current measurements, predict that the photon acquires a significant right-handed component, in particular due to the exchange of a heavy fermion in the penguin loop~\cite{Atwood:1997zr}.

This Letter presents a study of the radiative decay \BtoKpipig, previously observed at the \B factories with a measured branching fraction of $( 27.6 \pm 2.2 ) \times 10^{-6}$~\cite{Nishida:2002me,Aubert:2005xk,PDG2012}.
The inclusion of charge-conjugate processes is implied throughout.
Information about the photon polarization is obtained from the angular distribution of the photon direction with respect to the normal to the plane defined by the momenta of the three final-state hadrons in their center-of-mass frame.
The shape of this distribution, including the \emph{up-down asymmetry} between the number of events with the photon on either side of the plane, is determined.
%
This investigation is conceptually similar to the historical experiment that discovered parity violation by measuring the up-down asymmetry of the direction of a particle emitted in a weak decay with respect to an axial vector~\cite{PhysRev.105.1413}.
In \BtoKpipig decays, 
the up-down asymmetry is proportional to the photon polarization $\lambda_{\gamma}$~\cite{Gronau:2002rz,Kou:photon-polarization-k1:2010} and therefore measuring a value different from zero corresponds to demonstrating that the photon is polarized.
The currently limited knowledge of the structure of the \Kpipi mass spectrum, which includes interfering kaon resonances, prevents the translation of a measured asymmetry into an actual value for $\lambda_{\gamma}$.

%
The differential \BtoKpipig decay rate can be described in terms of $\theta$, defined in the rest frame of the final state hadrons as the angle between the direction opposite to the photon momentum $\vec{p}_{\gamma}$ and the normal $\vec{p}_{\pi,\text{slow}}\times\vec{p}_{\pi,\text{fast}}$ to the \Kpipi plane, where $\vec{p}_{\pi,\text{slow}}$ and $\vec{p}_{\pi,\text{fast}}$ correspond to the momenta of the lower and higher momentum pions, respectively.
Following the notation and developments of Ref.~\cite{Gronau:2002rz}, the differential decay rate of \BtoKpipig can be written as a fourth-order polynomial in $\cos\theta$ 
\begin{align}\label{eq:odd_even}
    \frac{\diff{\Gamma}}{\diff{s}\diff{s_{13}}\diff{s_{23}}\diff{\cos\theta}} \propto \sum_{i=0,2,4} a_i(s, s_{13}, s_{23})\cos^i\theta + \lambda_\gamma \sum_{j=1,3} a_j(s, s_{13}, s_{23})\cos^j\theta\,,
\end{align}
where $s_{ij}=(p_i+p_j)^2$ and $s=(p_1+p_2+p_3)^2$, 
and $p_1$, $p_2$ and $p_3$ are the four-momenta of the \pim, \pip and \Kp mesons, respectively.
The functions $a_k$ depend on the resonances present in the \Kpipi mass range of interest and their interferences.
The up-down asymmetry is defined as 
\begin{equation}\label{eq:updown}
\asym{} \equiv \frac{\int_0^1\diff{\cos\theta}\frac{\diff{\Gamma}}{\diff{\cos\theta}}-\int_{-1}^0\diff{\cos\theta}\frac{\diff{\Gamma}}{\diff{\cos\theta}}}{\int_{-1}^1\diff{\cos\theta}\frac{\diff{\Gamma}}{\diff{\cos\theta}}}\,,
\end{equation}
which is proportional to $\lambda_{\gamma}$.

The \lhcb detector~\cite{Alves:2008zz} is a single-arm forward
spectrometer covering the \mbox{pseudorapidity} range $2<\eta <5$,
designed for the study of particles containing \bquark or \cquark
quarks. The detector includes a high-precision tracking system
consisting of a silicon-strip vertex detector surrounding the $pp$
interaction region, a large-area silicon-strip detector located
upstream of a dipole magnet with a bending power of about
$4{\rm\,Tm}$, and three stations of silicon-strip detectors and straw
drift tubes placed downstream.
The combined tracking system provides a momentum measurement with
relative uncertainty that varies from 0.4\% at 5\gevc to 0.6\% at 100\gevc,
and impact parameter resolution of 20\mum for
tracks with a few \gevc of transverse momentum (\pt). Different types of charged hadrons are distinguished by information
from two ring-imaging Cherenkov detectors. 
Photon, electron and hadron candidates are identified by a calorimeter system consisting of scintillating-pad and preshower detectors, an electromagnetic
calorimeter and a hadronic calorimeter. 
The trigger 
consists of a hardware stage, based on information from the calorimeter and muon systems, followed by a software stage, which applies a full event
reconstruction.

Samples of simulated events are used to understand signal and backgrounds.
In the simulation, $pp$ collisions are generated using
\pythia~\cite{Sjostrand:2006za,*Sjostrand:2007gs} 
with a specific \lhcb
configuration~\cite{LHCb-PROC-2010-056}.  Decays of hadronic particles
are described by \evtgen~\cite{Lange:2001uf}, in which final state
radiation is generated using \photos~\cite{Golonka:2005pn}. The
interaction of the generated particles with the detector and its
response are implemented using the \geant
toolkit~\cite{Allison:2006ve, *Agostinelli:2002hh} as described in Ref.~\cite{LHCb-PROC-2011-006}.


This analysis is based on 
 the \lhcb data sample collected from $pp$ collisions at $7$ and $8$\,TeV center-of-mass energies in 2011 and 2012, respectively, corresponding to $3~\invfb$ of integrated luminosity.
The \BtoKpipig candidates are built from a photon candidate and a hadronic system reconstructed from a kaon and two oppositely charged pions satisfying particle identification requirements.
Each of the hadrons is required to have a minimum \pt of $0.5\,\gevc$ and at least one of them needs to have a \pt larger than $1.2\,\gevc$. 
The isolation of the \Kpipi vertex from other tracks in the event is ensured by requiring that the \chisq of the three-track vertex fit and the \chisq of all possible vertices that can be obtained by adding an extra track differ by more than $2$.
The \Kpipi mass is required to be in the $[1.1, 1.9]\,$\gevcc range.
High transverse energy ($> 3.0\,\gev$) photon candidates are constructed from energy depositions in the electromagnetic calorimeter. 
The absence of tracks pointing to the calorimeter is used to distinguish neutral from charged electromagnetic particles. 
A multivariate algorithm based on the energy cluster shape parameters is used to reject approximately $65\,\%$ of the \decay{\piz}{\g\g} background in which the two photons are reconstructed as a single cluster, while keeping about $95\,\%$ of the signal photons.
The \Bp candidate mass is required to be in the $[4.3, 6.9]\,$\gevcc range.
Backgrounds that are expected to peak in this mass range are suppressed by removing all candidates consistent with a \decay{\bar\Dz}{\Kp\pim\piz} or \decay{\rho^+}{\pip\piz} decay when the photon candidate is assumed to be a \piz.

A boosted decision tree~\cite{Breiman,AdaBoost} is used to further improve the separation between signal and background.
Its training is based on the following variables: the impact parameter \chisq of the \Bp meson and of each of the final state hadrons, defined as the difference between the \chisq of a primary vertex (PV) reconstructed with and without the considered particle; the cosine of the angle between the reconstructed \Bp momentum and the vector pointing from the PV to the \Bp decay vertex; the flight distance of the \Bp meson; and the \Kpipi vertex \chisq.

The mass distribution of the selected \BtoKpipig signal is modeled with a double-tailed Crystal Ball~\cite{CrystalBall} probability density function (\PDF), with power-law tails above and below the \B mass.
The four tail parameters are fixed from simulation; the width of the signal peak is fit separately for 2011 and 2012 data to account for differences in calorimeter calibration.
Combinatorial and partially reconstructed backgrounds are considered in the fit, the former modeled with an exponential \PDF, the latter described using an ARGUS \PDF~\cite{Albrecht1990278} convolved with a Gaussian function with the same width as the signal to account for the photon energy resolution.
The contribution to the partially reconstructed background from events with only one missing pion is considered separately.

\begin{figure}[t]
    \centering
    \includegraphics[width=0.49\textwidth]{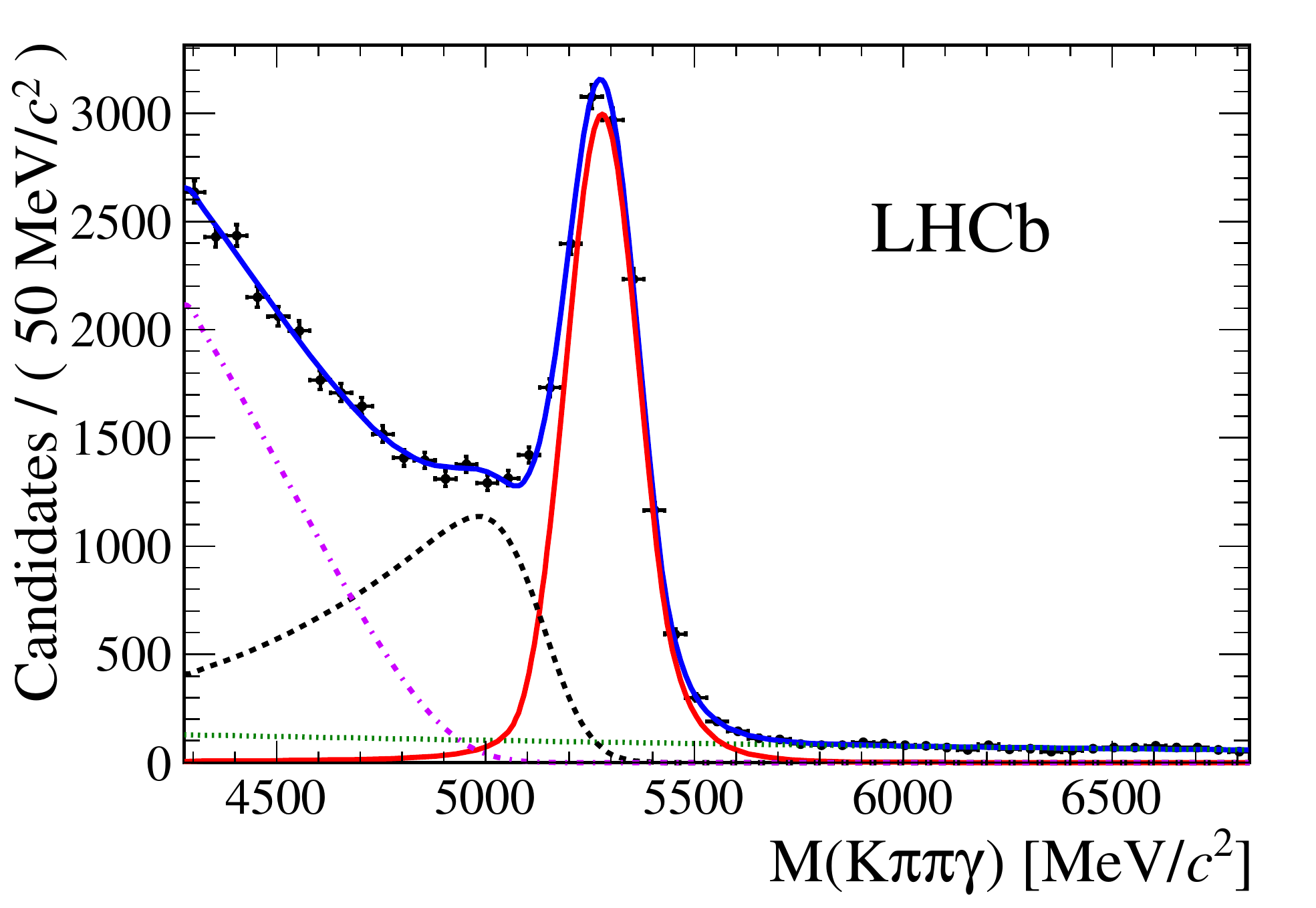}
    \caption{\small 
       Mass distribution of the selected \BtoKpipig candidates.
       The blue solid curve shows the fit results as the sum of the following components: signal  (red solid), combinatorial background (green dotted), missing pion background (black dashed) and other partially reconstructed backgrounds (purple dash-dotted).
        \label{fig:simulFits}}
\end{figure}

\begin{figure}[t]
    \centering
    \includegraphics[width=0.49\textwidth]{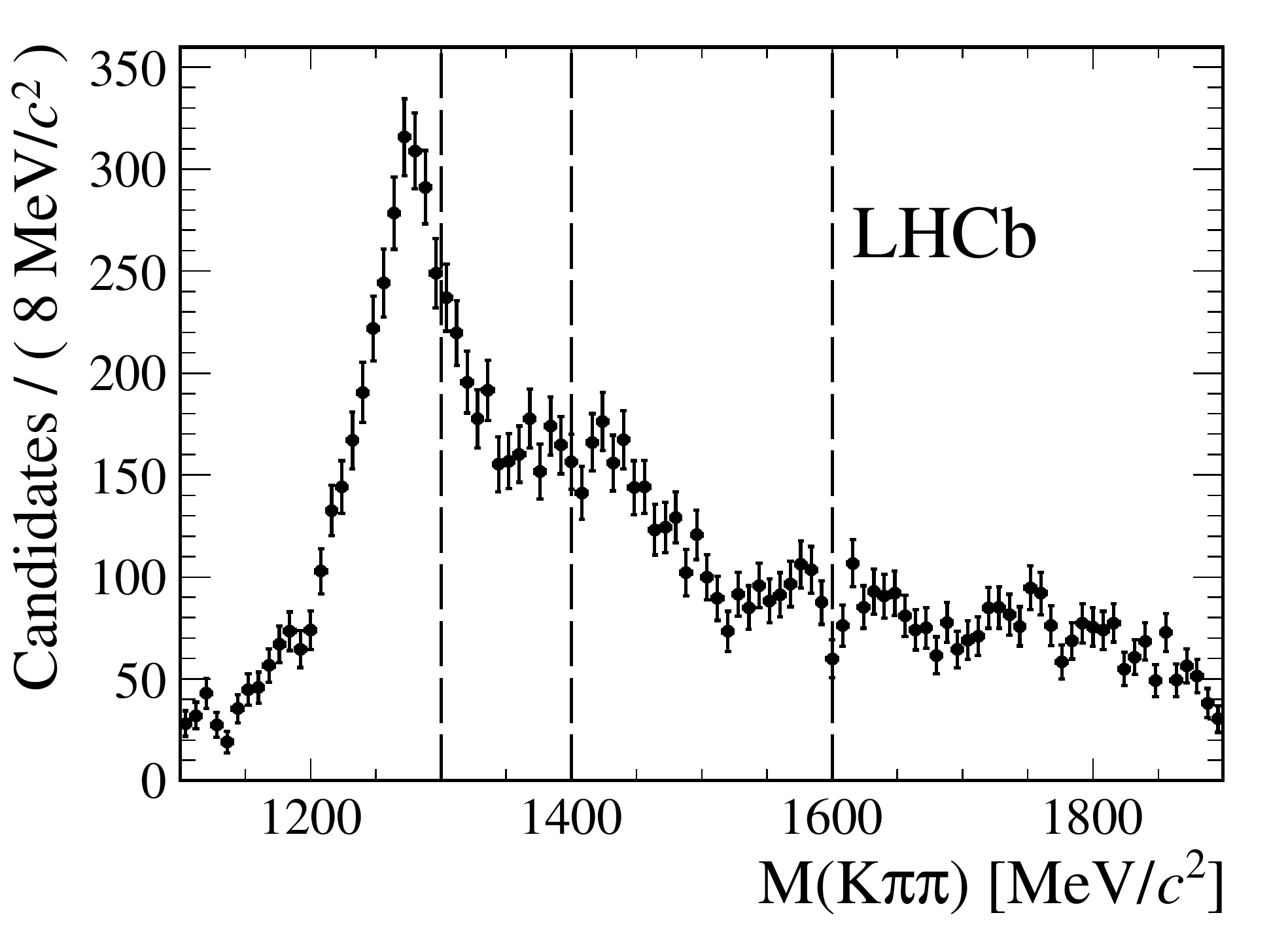}
  \caption{
    \small
    Background-subtracted \Kpipi mass distribution of the \BtoKpipig signal.
    The four intervals of interest, separated by dashed lines, are shown.}
  \label{fig:KpipiSplot}
\end{figure}

The fit of the mass distribution of the selected \BtoKpipig candidates (Fig.~\ref{fig:simulFits}) returns a total signal yield of $13\,876 \pm 153$ events, the largest sample recorded for this channel to date.
Figure~\ref{fig:KpipiSplot} shows the background-subtracted \Kpipi mass spectrum determined using the technique of Ref.~\cite{Pivk:2004ty}, after constraining the \B mass to its nominal value.
No peak other than that of the \Konep{1270} resonance can be clearly identified.
Many kaon resonances, with various masses, spins and angular momenta, are expected to contribute and interfere in the considered mass range~\cite{PDG2012}.

The contributions from single resonances cannot be isolated because of the complicated structure of the \Kpipi mass spectrum. 
The up-down asymmetry is thus studied inclusively in four intervals of \Kpipi mass.  
The $[1.4,1.6]\,\gevcc$ interval, studied in Ref.~\cite{Gronau:2002rz}, includes the \Konep{1400}, \Kstp{2}{1430} and \Kstp{}{1410} resonances 
 with small contributions from the upper tail of the \Konep{1270}.
At the time of the writing of Ref.~\cite{Gronau:2002rz}, the \Konep{1400} was believed to be the dominant $1^+$ resonance, so the \Konep{1270} was not considered.
However, subsequent experimental results~\cite{Yang:2004as} demonstrated that the \Konep{1270} is more prominent than the \Konep{1400}, hence the $[1.1,1.3]\,\gevcc$ interval is also studied here. 
The $[1.3,1.4]\,\gevcc$ mass interval, which contains the overlap region between the two $K_1$ resonances, and the $[1.6,1.9]\,\gevcc$ high mass interval, which includes spin-2 and spin-3 resonances, are also considered.

In each of the four \Kpipi mass intervals, a simultaneous fit to the \B-candidate mass spectra in bins of the photon angle is performed in order to determine the background-subtracted angular distribution; the previously described \PDF is used to model the mass spectrum in each bin, with all of the fit parameters being shared except for the yields.
Since the sign of the photon polarization depends on the sign of the electric charge of the \B candidate, the angular variable 
$\coshtheta\equiv\operatorname{charge}(\B)\costheta\,$ is used. 
The resulting background-subtracted \coshtheta distribution, corrected for the selection acceptance and normalized to the inverse of the bin width, is fit with a fourth-order polynomial function normalized to unit area,
\begin{equation}\label{eq:legendre}
    f(\coshtheta; c_0\!=\!0.5, c_1, c_2, c_3, c_4) = \sum_{i=0}^4 c_{i}L_i(\coshtheta)\,,
\end{equation}
where $L_i(x)$ is the Legendre polynomial of order $i$ and $c_i$ is the corresponding coefficient.
Using Eqs.~\ref{eq:odd_even} and~\ref{eq:legendre} the up-down asymmetry defined in Eq.~\ref{eq:updown} can be expressed as 
\begin{equation}\label{eq:aUpDown}
    \asym{} = c_1 - \frac{c_3}{4} \,\,.
\end{equation}
As a cross-check, the up-down asymmetry in each mass interval is also determined with a counting method, rather than an angular fit, as well as considering separately the \Bp and \Bm candidates. 
All these checks yield compatible results.

The results obtained from a \chisq fit of the normalized binned angular distribution, performed taking into account the full covariance matrix of the bin contents and all of the systematic uncertainties, are summarized in Table~\ref{tab:angular_Fit}.
These systematic uncertainties account for the effect of choosing a different fit model, the impact of the limited size of the simulated samples on the fixed parameters, and the possibility of some events migrating from a bin to its neighbor because of the detector resolution, which gives the dominant contribution.
The systematic uncertainty associated with the fit model is determined by performing the mass fit using several alternative PDFs, while the other two are estimated with simulated pseudoexperiments.
Such uncertainties, despite being of the same size as the statistical uncertainty, do not substantially affect the fit results since they are strongly correlated across all angular bins.

\begin{figure*}[t]
    \centering
    \captionsetup[subfigure]{labelformat=empty}
    \includegraphics[width=0.4\textwidth]{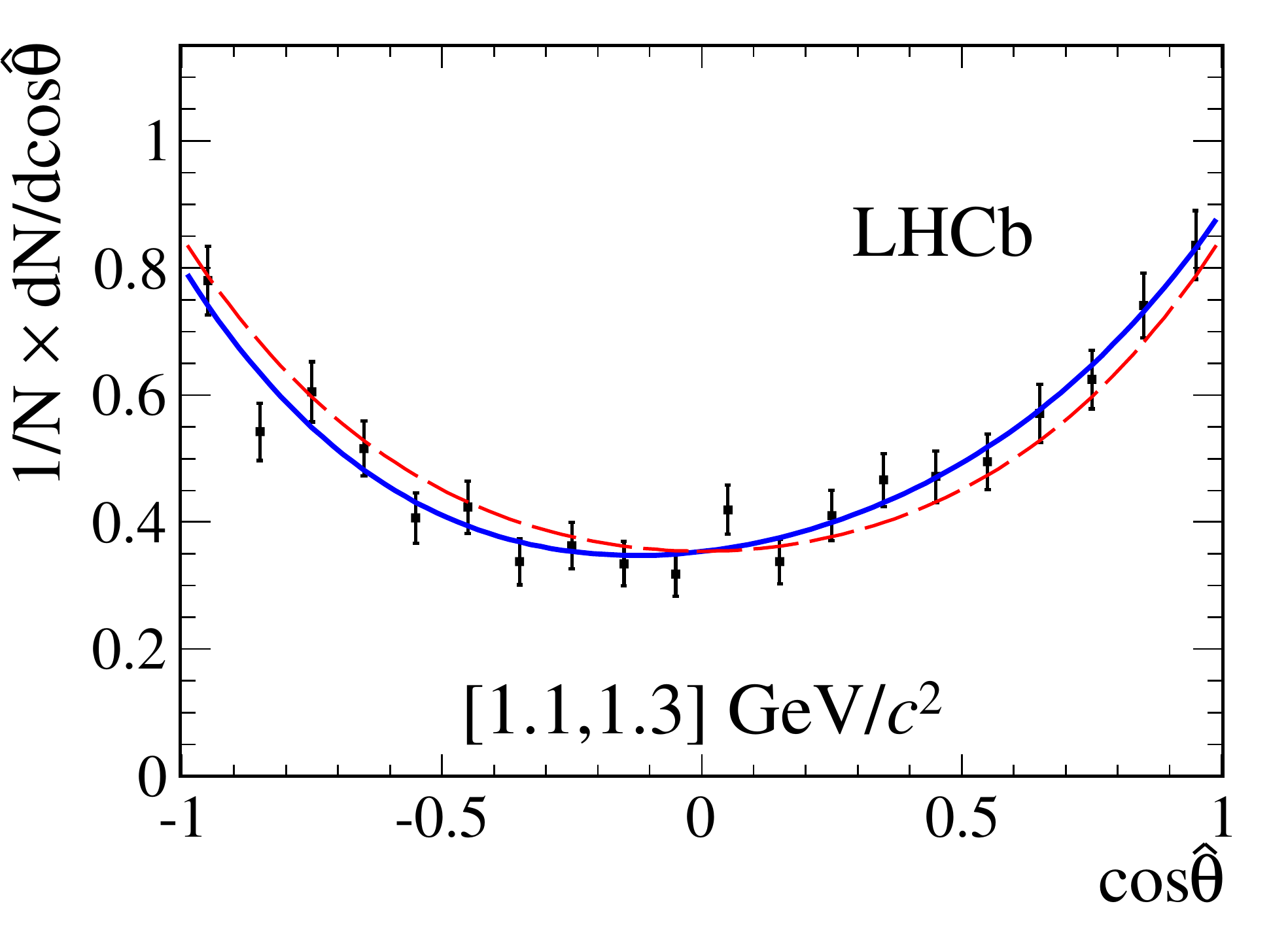}
    \includegraphics[width=0.4\textwidth]{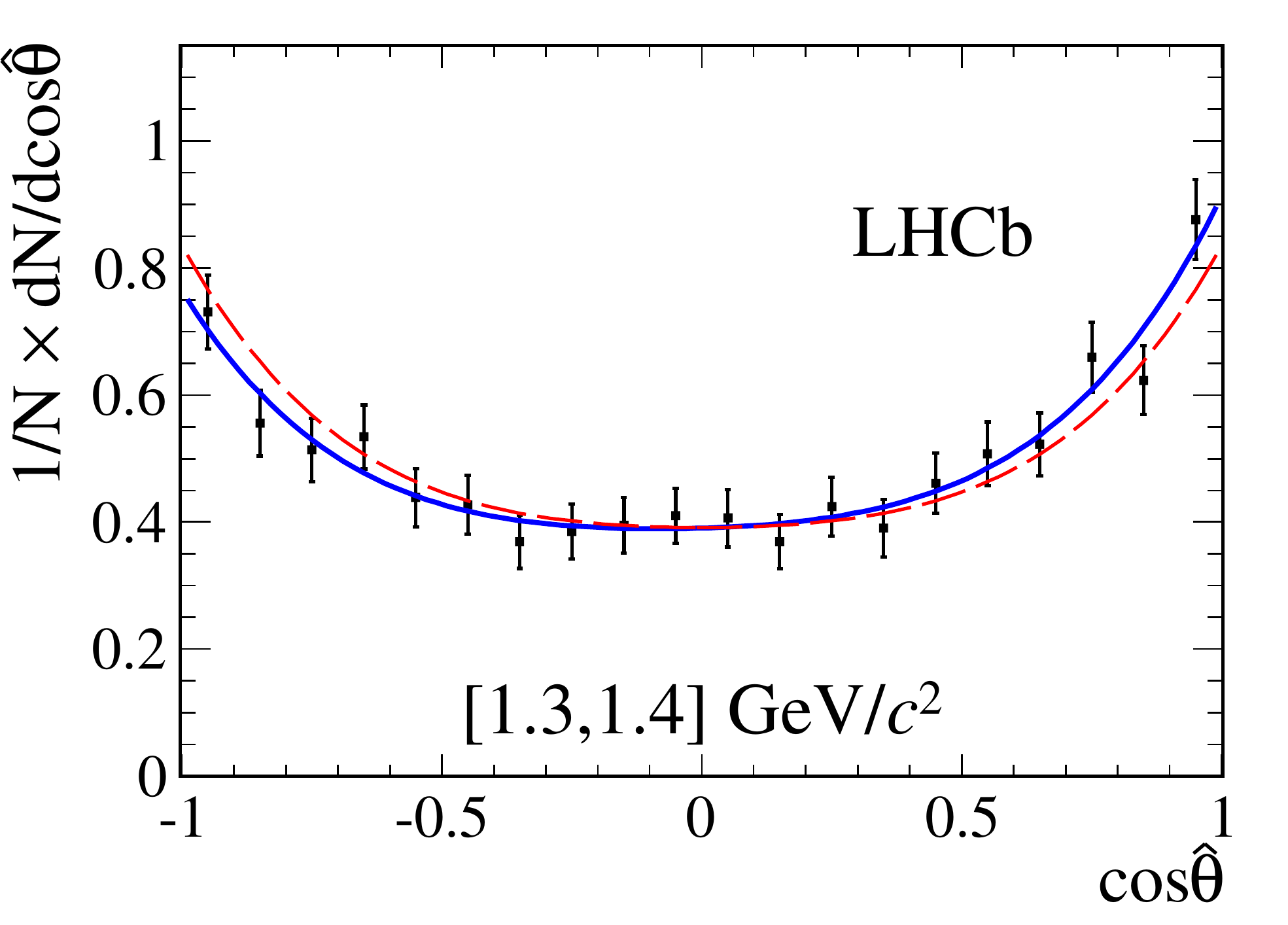}\\
    \includegraphics[width=0.4\textwidth]{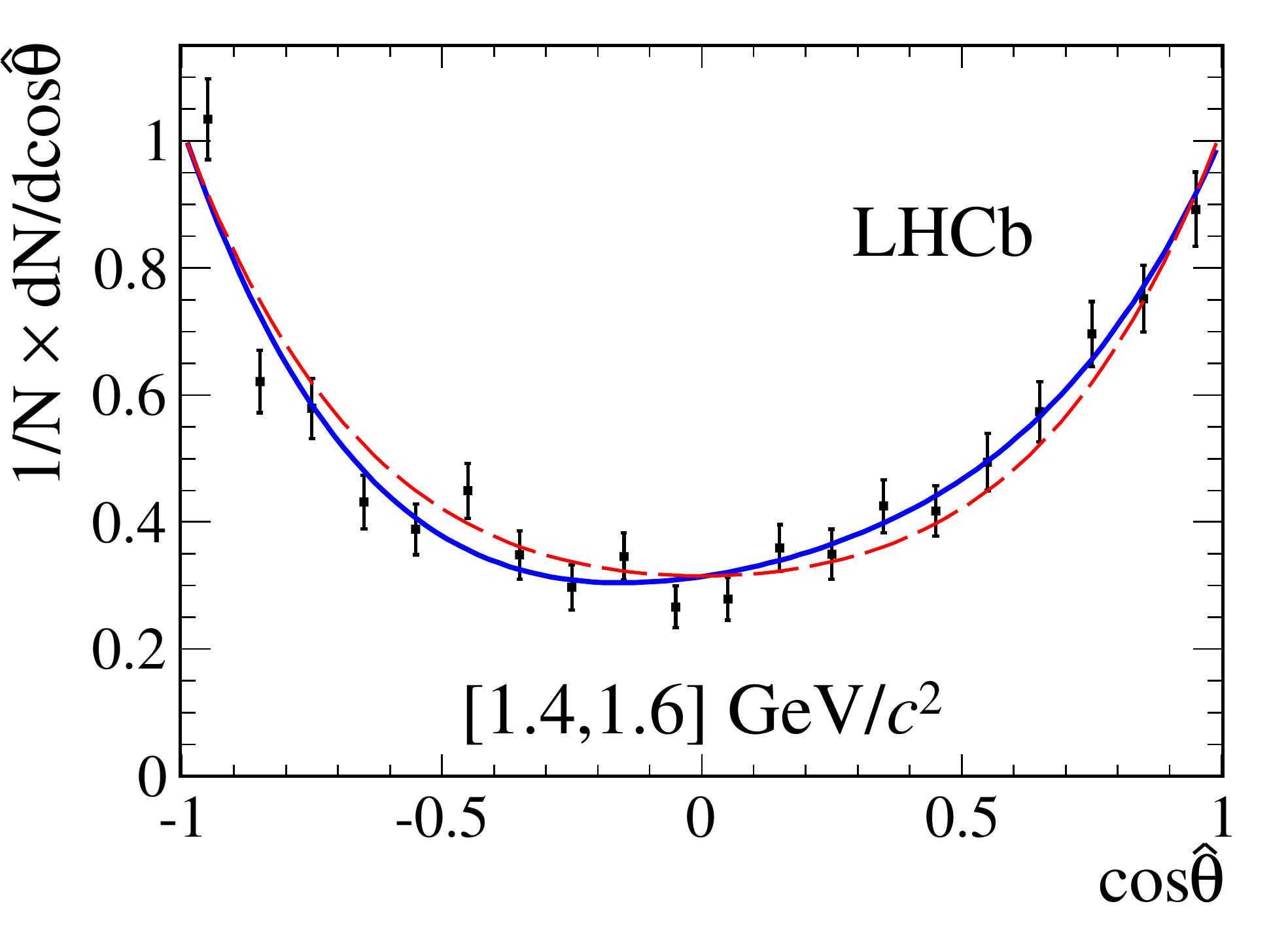}
    \includegraphics[width=0.4\textwidth]{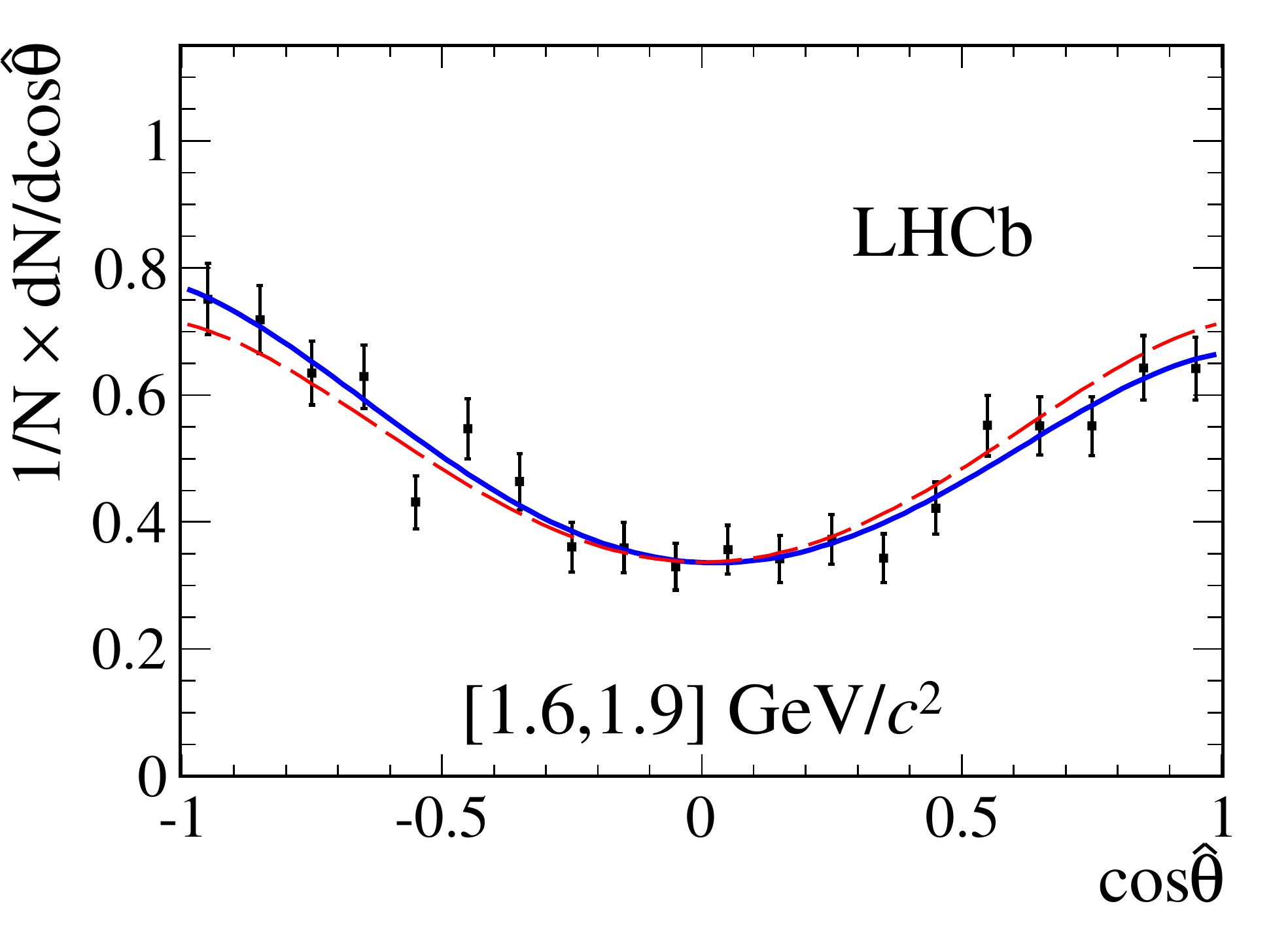}
    \caption{\small 
        Distributions of \coshtheta for \BtoKpipig signal in four intervals of \Kpipi mass. 
        The solid blue (dashed red) curves are the result of fits allowing all (only even) Legendre components up to the fourth power. 
        }
        \label{fig:cosTheta_Fit}
\end{figure*}

The fitted distributions in the four \Kpipi mass intervals of interest are shown in Fig.~\ref{fig:cosTheta_Fit}.
In order to illustrate the effect of the up-down asymmetry, the results of another fit imposing $c_1=c_3=0$, hence forbidding the terms that carry the $\lambda_{\gamma}$ dependence, are overlaid for comparison. 

The combined significance of the observed up-down asymmetries is determined from a \chisq test where the null hypothesis is defined as $\lambda_{\gamma}=0$, implying that the up-down asymmetry is expected to be zero in each mass interval. 
The corresponding \chisq distribution has four degrees of freedom, and the observed value corresponds to a p-value of $1.7 \times 10^{-7}$.
This translates into a $5.2\,\sigma$ significance for nonzero up-down asymmetry. 
Up-down asymmetries can be computed also for an alternative definition of the photon angle, obtained using the normal $\vec{p}_{\pim}\times\vec{p}_{\pip}$ instead of $\vec{p}_{\pi,\text{slow}}\times\vec{p}_{\pi,\text{fast}}$. 
The obtained values, along with the relative fit coefficients, are listed in Table~\ref{tab:angularFit_theta}.
\begin{table}[ht]
  \caption{
    \small 
    Legendre coefficients obtained from fits to the normalized background-subtracted \coshtheta distribution in the four \Kpipi mass intervals of interest.
    The up-down asymmetries are obtained from Eq.~\ref{eq:aUpDown}.
    The quoted uncertainties contain statistical and systematic contributions.
    The \Kpipi mass ranges are indicated in\,\gevcc and all the parameters are expressed in units of $10^{-2}$. 
    The covariance matrices are given in the supplementary material.}
   \centering
\begin{tabular}{@{}c@{  }r@{}c@{}l@{   }r@{}c@{}l@{   }r@{}c@{}l@{   }r@{}c@{}l}
    \toprule
             & \multicolumn{3}{c}{$[1.1,1.3]$}    & \multicolumn{3}{c}{$[1.3,1.4]$}    & \multicolumn{3}{c}{$[1.4,1.6]$}    & \multicolumn{3}{c}{$[1.6,1.9]$} \\
  \midrule
  $c_1$    & $6.3 $  & $\pm$ & $ 1.7$ & $5.4 $  & $\pm$ & $ 2.0$ & $4.3 $  & $\pm$ & $ 1.9$ & $-4.6 $ & $\pm$ & $ 1.8$   \\
  $c_2$    & $31.6 $ & $\pm$ & $ 2.2$ & $27.0 $ & $\pm$ & $ 2.6$ & $43.1 $ & $\pm$ & $ 2.3$ & $28.0 $ & $\pm$ & $ 2.3$      \\
  $c_3$    & $-2.1 $ & $\pm$ & $ 2.6$ & $2.0 $  & $\pm$ & $ 3.1$ & $-5.2 $ & $\pm$ & $ 2.8$ & $-0.6 $ & $\pm$ & $ 2.7$ \\
  $c_4$    & $3.0 $  & $\pm$ & $ 3.0$ & $6.8 $  & $\pm$ & $ 3.6$ & $8.1 $  & $\pm$ & $ 3.1$ & $-6.2 $ & $\pm$ & $ 3.2$   \\
  \midrule
  \asym{}  & $6.9 $  & $\pm$ & $ 1.7$ & $4.9 $  & $\pm$ & $ 2.0$ & $5.6 $  & $\pm$ & $ 1.8$ & $-4.5 $ & $\pm$ & $ 1.9$     \\
  \bottomrule
\end{tabular}
\label{tab:angular_Fit}
\end{table}
\begin{table}[ht]
  \caption{
    \small 
    Legendre coefficients obtained from fits to the normalized background-subtracted \coshtheta distribution, using the alternative normal $\vec{p}_{\pim}\times\vec{p}_{\pip}$ for defining the direction of the photon, in the four \Kpipi mass intervals of interest.
    The up-down asymmetries are obtained from Eq.~\ref{eq:aUpDown}.
    The quoted uncertainties contain statistical and systematic contributions.
    The \Kpipi mass ranges are indicated in\,\gevcc and all the parameters are expressed in units of $10^{-2}$. 
    The covariance matrices are given in the supplementary material.}
   \centering
\begin{tabular}{@{}c@{ }r@{}c@{}l@{   }r@{}c@{}l@{   }r@{}c@{}l@{   }r@{}c@{}l}
    \toprule
             & \multicolumn{3}{c}{$[1.1,1.3]$}    & \multicolumn{3}{c}{$[1.3,1.4]$}    & \multicolumn{3}{c}{$[1.4,1.6]$}    & \multicolumn{3}{c}{$[1.6,1.9]$} \\
  \midrule
  $c_1^\prime$       & $-0.9 $&$\pm$&$ 1.7$ & $7.4 $&$\pm$&$ 2.0$  & $5.3 $&$\pm$&$ 1.9$  & $-3.4 $&$\pm$&$ 1.8$   \\
  $c_2^\prime$       & $31.6 $&$\pm$&$ 2.2$ & $27.4 $&$\pm$&$ 2.6$ & $43.6 $&$\pm$&$ 2.3$ & $27.8 $&$\pm$&$ 2.3$      \\
  $c_3^\prime$       & $0.8 $&$\pm$&$ 2.6$  & $0.8 $&$\pm$&$ 3.1$  & $-4.4 $&$\pm$&$ 2.8$ & $2.3 $&$\pm$&$ 2.7$ \\
  $c_4^\prime$       & $3.4 $&$\pm$&$ 3.0$  & $7.0 $&$\pm$&$ 3.6$  & $8.0 $&$\pm$&$ 3.1$  & $-6.6 $&$\pm$&$ 3.2$   \\
  \midrule
  $\mathcal{A}^{\prime}_{\text{ud}}$     & $-1.1 $&$\pm$&$ 1.7$ & $7.2 $&$\pm$&$ 2.0$  & $6.4 $&$\pm$&$ 1.8$  & $-3.9 $&$\pm$&$ 1.9$     \\
  \bottomrule
\end{tabular}
\label{tab:angularFit_theta}
\end{table}

To summarize, 
a study of the inclusive flavor-changing neutral current radiative \BtoKpipig decay, with the \Kpipi mass in the $[1.1,1.9]\,\gevcc$ range, is performed on a data sample corresponding to an integrated luminosity of $3\,\invfb$ collected in $pp$ collisions at $7$ and $8$\,TeV center-of-mass energies by the \lhcb detector.
A total of $13\,876 \pm 153$ signal events is observed.
The shape of the angular distribution of the photon with respect to the plane defined by the three final-state hadrons in their rest frame is determined in four intervals of interest in the \Kpipi mass spectrum.
The up-down asymmetry, which is proportional to the photon polarization, is measured for the first time for each of these \Kpipi mass intervals.
The first observation of a parity-violating photon polarization different from zero at the $5.2\,\sigma$ significance level in \btosgam transitions is reported.
The shape of the photon angular distribution in each bin, as well as the values for the up-down asymmetry, may be used, if theoretical predictions become available, to determine for the first time a value for the photon polarization, and thus constrain the effects of physics beyond the SM in the \btosgam sector.

\clearpage
\section*{Acknowledgements}

\noindent We express our gratitude to our colleagues in the CERN
accelerator departments for the excellent performance of the LHC. We
thank the technical and administrative staff at the LHCb
institutes. We acknowledge support from CERN and from the national
agencies: CAPES, CNPq, FAPERJ and FINEP (Brazil); NSFC (China);
CNRS/IN2P3 and Region Auvergne (France); BMBF, DFG, HGF and MPG
(Germany); SFI (Ireland); INFN (Italy); FOM and NWO (The Netherlands);
SCSR (Poland); MEN/IFA (Romania); MinES, Rosatom, RFBR and NRC
``Kurchatov Institute'' (Russia); MinECo, XuntaGal and GENCAT (Spain);
SNSF and SER (Switzerland); NAS Ukraine (Ukraine); STFC (United
Kingdom); NSF (USA). We also acknowledge the support received from the
ERC under FP7. The Tier1 computing centres are supported by IN2P3
(France), KIT and BMBF (Germany), INFN (Italy), NWO and SURF (The
Netherlands), PIC (Spain), GridPP (United Kingdom).
We are indebted towards the communities behind the multiple open source software packages we depend on.
We are also thankful for the computing resources and the access to software R\&D tools provided by Yandex LLC (Russia).

\mciteErrorOnUnknownfalse

\addcontentsline{toc}{section}{References}
\setboolean{inbibliography}{true}
\bibliographystyle{LHCb}
\bibliography{main}

\clearpage


\section*{Supplementary material}
\label{sec:Supplementary-App}

The covariance matrices obtained from the fit described in the Letter for both photon angle definitions are shown in Tables~\ref{tab:angular_flip_cov} and~\ref{tab:angular_noflip_cov}.

\makeatletter
  \def\env@matrix{\hskip -\arraycolsep
  \let\@ifnextchar\new@ifnextchar
  \array{*\c@MaxMatrixCols r}}
\makeatother
  
\begin{table}[ht]
  \caption{
    \small 
    Covariance matrices (in units of $10^{-3}$) for the fitted values of $c_1$, $c_2$, $c_3$ and $c_4$ of Table~\ref{tab:angular_Fit}, for the four $\Kpipi$ mass intervals.
}
   \centering
    \captionsetup[subfloat]{labelformat=empty}
    \subfloat[][{$[1.1,1.3]\,\gevcc$}]{$\begin{pmatrix}
{\phantom+}0.31 & \phantom{+0.01} & \phantom{+0.09} & \phantom{-0.01} \\
0.01            & 0.47            &                 &   \\
0.09            & 0.03            & 0.68            &   \\
-0.01           & 0.16            & 0.02            & 0.92 \\
\end{pmatrix}$} \hspace{1cm}
    \subfloat[][{$[1.3,1.4]\,\gevcc$}]{$\begin{pmatrix}
0.41 & \phantom{+0.02} & \phantom{+0.12} & \phantom{+0.00}  \\
0.02 & 0.66            &                 &   \\
0.12 & 0.04            & 0.93            &  \\
0.00 & 0.20            & 0.04            & 1.27   \\
    \end{pmatrix}$}\\
    \subfloat[][{$[1.4,1.6]\,\gevcc$}]{$\begin{pmatrix}
0.35  & \phantom{-0.01} &       & \phantom{-0.03} \\
-0.01 & 0.52            &       & \\
0.14  & 0.00            & 0.76  & \\
-0.03 & 0.23            & -0.01 & 0.99 \\
    \end{pmatrix}$} \hspace{1cm}
    \subfloat[][{$[1.6,1.9]\,\gevcc$}]{$\begin{pmatrix}
0.34  & \phantom{-0.02} & \phantom{0.08}  & \phantom{-0.02} \\
-0.02 & 0.51  &  &    \\
0.08  & -0.04 & 0.75  &   \\
-0.02 & 0.15  & -0.04 & 1.01  \\
    \end{pmatrix}$}
\label{tab:angular_flip_cov}
\end{table}

\begin{table}[htpb]
  \caption{
    \small 
    Covariance matrices (in units of $10^{-3}$) for the fitted values of $c^{\prime}_1$, $c^{\prime}_2$, $c^{\prime}_3$ and $c^{\prime}_4$ of Table~\ref{tab:angularFit_theta}, for the four $\Kpipi$ mass intervals.
}
   \centering
    \captionsetup[subfloat]{labelformat=empty}
    \subfloat[][{$[1.1,1.3]\,\gevcc$}]{$\begin{pmatrix}
0.30 & \phantom{+0.00} & \phantom{+0.09} & \phantom{+0.02}  \\
0.00 & 0.47            &             &   \\
0.09 & 0.02            & 0.68            &   \\
0.02 & 0.16            & 0.02            & 0.92 \\
    \end{pmatrix}$} \hspace{1cm}
    \subfloat[][{$[1.3,1.4]\,\gevcc$}]{$\begin{pmatrix}
0.41 & \phantom{+0.03} & \phantom{+0.12} & \phantom{+0.01}  \\
0.03 & 0.66            &             &    \\
0.12 & 0.07            & 0.93            &   \\
0.01 & 0.20            & 0.10            & 1.27  \\
    \end{pmatrix}$}\\
    \subfloat[][{$[1.4,1.6]\,\gevcc$}]{$\begin{pmatrix}
0.35 & \phantom{+0.01} & \phantom{+0.14} & \phantom{+0.00} \\
0.01 & 0.53            &             &  \\
0.14 & 0.05            & 0.76            &  \\
0.00 & 0.24            & 0.03            & 0.99 \\
    \end{pmatrix}$} \hspace{1cm}
    \subfloat[][{$[1.6,1.9]\,\gevcc$}]{$\begin{pmatrix}
0.34 & \phantom{+0.00} & \phantom{+0.08} & \phantom{+0.02} \\
0.00 & 0.51            &             &  \\
0.08 & 0.00            & 0.75            &  \\
0.02 & 0.15            & -0.01           & 1.01 \\
    \end{pmatrix}$}
\label{tab:angular_noflip_cov}
\end{table}

\clearpage



\end{document}